\documentclass[11pt]{article}

 \newcommand{\be}{\begin{equation}}
 \newcommand{\ee}{\end{equation}}
 \newcommand{\ba}{\begin{eqnarray}}
 \newcommand{\ea}{\end{eqnarray}}
 \newcommand{\bl}{\begin{equation}\begin{array}{ll}}
 \newcommand{\el}{\end{array}\end{equation}}
 \newcommand{\bll}{\begin{equation}\begin{array}{lll}}
 \newcommand{\bdm}{\begin{displaymath}}
 \newcommand{\edm}{\end{displaymath}}
 \def\p{\partial}
 \def\f{\varphi}
 \def\ve{\varepsilon}
 
\def\lim{\rightarrow}
\def\half{\frac{1}{2}}

\def\be{\begin{equation}}
\def\ee{\end{equation}}
\def\bea{\begin{eqnarray}}
\def\eea{\end{eqnarray}}

\def\dif{\partial}

\def\bu{\bar{u}}

\def\bv{\bar{v}}

\begin{document}
\raggedbottom

\title{A New Integrable Model of (1+1)-Dimensional \\
          Dilaton  Gravity Coupled  to Toda Matter}

\author{A.T.~Filippov\thanks{Alexandre.Filippov@jinr.ru}~ \\
{\small \it {$^+$ Joint Institute for Nuclear Research, Dubna, Moscow
Region RU-141980} }}

\maketitle

\begin{abstract}

A new class of integrable two-dimensional dilaton gravity theories, in
  which scalar matter fields satisfy the Toda equations, is proposed. The
simplest case of the  Toda system is considered in some detail, and on
this example we outline how the general solution can be obtained.
       We also
  demonstrate how the wave-like solutions of the general Toda systems can
be simply derived.
       In the dilaton gravity theory, these solutions describe
nonlinear waves coupled to gravity.  A special attention is paid to making
the analytic structure of the solutions of the Toda equations as simple
and transparent as possible, with the aim to apply the idea of the
separation of variables to non-integrable theories.

\end{abstract}

\section{Introduction}

The theories of (1+1)-dimensional dilaton gravity coupled to scalar matter
fields are known to be reliable models for some aspects of
higher-dimensional black holes, cosmological models, and waves. The
  connection between higher and lower dimensions was demonstrated in
different contexts of gravity and string theory and, in several cases, has
  allowed finding the general solution or special classes of solutions
in high-dimensional theories
 \footnote{See, e.g., \cite{BZ}-\cite{Venezia} for a more detailed discussion
 of this connection, references, and solution of some integrable two-dimensional
 and one-dimensional models of dilaton gravity. }. A generic example is the
spherically symmetric gravity coupled to Abelian gauge fields and massless
scalar matter fields. It exactly reduces to a (1+1)-dimensional dilaton
gravity and can be explicitly solved if the scalar fields are constants
  independent of the coordinates. These solutions can describe interesting
physical objects -- spherical static black holes and simplest cosmologies.
However, when the scalar matter fields, which presumably play a
significant cosmological role, are nontrivial, not many exact analytical
solutions of high-dimensional theories are known\footnote{See, e.g.,
         \cite{CGHS}, \cite{NKS}, \cite{ATF4}, \cite{VDA1}-\cite{ATF5};
         a review and further references can be found in \cite{Strobl},
 \cite{Kummer} and \cite{ATF5}}.
  Correspondingly, the two-dimensional models of dilaton gravity that
  nontrivially couple to scalar matter are usually not integrable.

   To obtain integrable models of this sort one usually has to make
 serious approximations, in other words, to deform the original
two-dimensional model obtained by direct dimensional reductions of
  realistic higher-dimensional theories\footnote{ We note that several
important four-dimensional space-times with symmetries defined by two
       commuting Killing vectors may also be described by two-dimensional
       models of dilaton gravity coupled to scalar matter. For example,
  cylindrical gravitational waves can be described by a (1+1)-dimensional
  dilaton gravity   coupled to one scalar field
  \cite{Einstein}-\cite{Chandra1}, \cite{ATF3}.
 The stationary axially symmetric pure gravity (\cite{Ernst}, \cite{NKS})
  is equivalent to a (0+2)-dimensional dilaton gravity coupled to one scalar
field. Similar but more general dilaton gravity models were also obtained
in string theory. Some of them can be solved by using modern mathematical
methods developed in the soliton theory (see e.g. \cite{BZ},
\cite{Maison}, \cite{NKS}, \cite{Alekseev}). }.
 Nevertheless, the deformed models can qualitatively describe certain
 physically interesting solutions of higher-dimensional gravity or
supergravity theories related to the low-energy limit of superstring
theories.

In our previous work (see, e.g., \cite{ATF2} - \cite{ATF5} and references
therein) we constructed and studied some explicitly integrable models
 based on the Liouville equation.
 Recently, we attempted to find solutions of some realistic
     two-dimensional dilaton gravity models (derived from higher-dimensional
      gravity theories by dimensional reduction) using a generalized
 separation of variables introduced in \cite{W1}, \cite{ATF3}.
 These attempts showed that seemingly
natural ansatzes for the structure of the separation, which proved a
  success in previously studied integrable models, do not give
interesting enough solutions (`zero' approximation of a perturbation
theory) in realistic nonintegrable models. Thus an investigation of more
complex dilaton gravity models, which are based on the two dimensional
 Toda  chains, was initiated. Here we briefly present some results of this
 investigation. In particular, we propose a simplest class of models, which
  can be explicitly solved in terms of the solutions of the Toda equations.
 These solutions as well as their moduli space will be presented in a very
 simple and convenient form that allows for a simple description of
 the analytic and asymptotic properties of the solutions.
 At the same time, this
 representation is extremely convenient for their reductions to the wave
 solutions that include static and cosmological ones. This construction
 essentially and naturally generalizes the previous results,
 \cite{ATF2} - \cite{ATF5}, and shows that essentially more complex structures
 of the separation of variables should be employed in realistic theories of
 gravity.

\section{General model of (1+1)-dimensional dilaton gravity \\
  minimally coupled to scalar matter fields.}

The effective Lagrangian of the (1+1)-dimensional dilaton gravity coupled
to scalar fields $\psi_n$ obtainable by dimensional reductions of a
higher-dimensional spherically symmetric (super)gravity can usually be
(locally) transformed to the form (see \cite{ATF2} - \cite{ATF5} for a
detailed motivation and specific examples):
\be
 {\cal L} = \sqrt{-g}\left[ \f R(g) + V(\f,\psi) +
 \sum_{mn} Z_{mn}(\f,\psi)\, g^{ij} \, {\dif}_i \psi_m \, {\dif}_j \psi_n \right] \, .
\label{7}
\ee
Here $g_{ij}(x^0,x^1)$ is the (1+1)-dimensional metric with the signature
 (-1,1), $g \equiv {\rm det}|g_{ij}|$, and $R$ is the Ricci curvature of
 the two-dimensional space-time,
\be
ds^2=g_{ij}\, dx^i \, dx^j \, , \,\,\,\,\,\, i,j = 0,1 \, .
\label{4}
\ee
The effective potentials $V$ and $Z_{mn}$ depend on the dilaton
 $\f(x^0,x^1)$ and on $N-2$ scalar fields $\psi_n(x^0,x^1)$
 (we note that the matrix $Z_{mn}$ should be negative definite
 to exclude the so called `phantom' fields).
 They may depend on other parameters characterizing the parent
higher-dimensional theory (e.g., on charges introduced in solving the
equations for the Abelian fields).
         Here we consider the `minimal'
 kinetic terms with the diagonal and constant $Z$-potentials,
 $Z_{mn}(\f, \psi) = \delta_{mn} Z_n$.  This approximation
 excludes the important class of the sigma - model - like
 scalar matter discussed, e.g., in \cite{Venezia}; such models can be
 integrable if $V \equiv 0$ and $Z_{mn}(\f,\psi)$ satisfy certain rather
 stringent conditions.
 In (\ref{7}) we also used the Weyl transformation to eliminate
 the gradient term for the dilaton.

          To simplify derivations, we write the equations of motion in the
light-cone metric,
  $$ds^2 = -4f(u, v) \, du \, dv \, .$$
 By first varying the Lagrangian in generic coordinates and then passing
 to the light-cone coordinates we obtain the equations of motion
  ($Z_n$ are constants!)
 \be
 \p_u \p_v \f+f\, V(\f,\psi) = 0, \label{F.15}
 \ee
  \be
  f \p_i ({{\p_i \f} / f }) \, = \sum Z_n (\p_i \psi_n)^2\, , \,\,\,\,\,\,\,\,\,
 i=u,v \, .
 \label{F.17}
 \ee
 \be
 2Z_n \p_u \p_v \psi_n + f V_{\psi_n}(\f,\psi)= 0 \, ,
 \label{F.16}
 \ee
 \be
 \p_u\p_v\ln |f| + f V_{\f}(\f,\psi) = 0 \, ,
 \label{F.18}
 \ee
 where $V_{\f}= \p_{\f} V$, $V_{\psi_n} = \p_{\psi_n} V$.
 These equations  are not independent. Actually, (\ref{F.18}) follows
from (\ref{F.15}) $-$  (\ref{F.16}). Alternatively, if  (\ref{F.15}),
(\ref{F.17}), and (\ref{F.18}) are satisfied, one of the equations
(\ref{F.16}) is also satisfied.

The higher-dimensional origin of the Lagrangian (\ref{7}) suggests that
the potential is the sum of the exponentials of linear combinations of
 the scalar fields, $q_n^{(0)}$, and of the dilaton $\f$~
       \footnote{Actually, the potential $V$ usually contains terms
       non exponentially depending on $\f$ (e.g., linear in $\f$),
       and then the exponentiation of $\f$ is only an approximation,
       see the discussion in \cite{ATF5}.}.
     In our previous work \cite{ATF5}
 we studied the constrained Liouville model, in which
the system of the equations of motion (\ref{F.15}), (\ref{F.16}) and
(\ref{F.18}) is equivalent to the system of the independent Liouville
equations for the
 linear combinations of the fields $q_n \equiv F + q_n^{(0)}$, where
 $F \equiv \ln|f|$. The easily derived solutions of these equations should
 satisfy  the constraints (\ref{F.17}), which was the most difficult part of
 the problem. The solution of the whole problem revealed an interesting
 structure of the moduli space of the solutions that allowed us to easily
 identify static, cosmological and wave-like solutions and effectively
 embed these essentially one-dimensional (in some broad sense) solutions
 into the set of all two-dimensional solutions and study their analytic
 and asymptotic properties.

 Here we propose a natural generalization of the Liouville model to the
 model in which the fields are described by the Toda equations (or by
 nonintegrable deformations of them). To demonstrate that the model shares
 many properties with the Liouville one and to simplify a transition from the
          integrable models to nonintegrable theories we suggest a different
 representation of the Toda solutions, which is not directly related to
 their group - theoretical background.

 Consider the theory defined by the Lagrangian (\ref{7})
 with the potential:
\be
 V = \sum_{n=1}^N 2g_n \exp{q_n^{(0)}} \, , \,\,\,\;\;\;\;\, Z_n =-1 \ ,
\label{8}
\ee
 where
\be
q_n^{(0)} \equiv a_n \f + \sum_{m=3}^{N} \psi_m a_{mn} \, .
 \label{9}
\ee
In what follows we also use
 \be
 q_n \equiv  F + q_n^{(0)} \equiv \sum_{m=1}^{N} \psi_m a_{mn} \, ,
 \label{9a}
 \ee
where $\psi_1 + \psi_2 \equiv \ln{|f|} \equiv F$ ($f \equiv \varepsilon
e^F$, $\varepsilon = \pm 1$), $\psi_1 - \psi_2 \equiv \f$ and hence
$a_{1n} = 1 + a_n$, $a_{2n} = 1 - a_n$.

 Rewriting the equations of motion in terms of $\psi_n$,
 we find that  Eqs.~(\ref{F.15}) - (\ref{F.18}) are equivalent to
 $N$ equations of motion for $N$ functions $\psi_n$,
\be
 \p_u \p_v \psi_n  =
  \varepsilon  \sum_{m=1}^{N} \epsilon_n a_{nm} g_m e^{q_m} \, ; \,\,\,\,
  \epsilon_1 = -1, \,\,\, \epsilon_n = +1, \,\, {\rm if} \,\, n \geq 2 \, ,
\label{10}
\ee
and two constraints,
\be
 \p_i^2 \f \equiv  \p_i^2 (\psi_1 - \psi_2)  =
 -\sum_{n=1}^N  \epsilon_n  (\p_i \psi_n)^2 , \,\,\,\,\,\,\, i=u,v \, .
 \label{11}
\ee
 With arbitrary parameters $a_{mn}$, these equations of motion are not
integrable.
 But as proposed in \cite{A2} - \cite{ATF1}, \cite{ATF2} \cite{ATF5},
 Eqs.(\ref{10}) are integrable and constraints (\ref{11}) can be solved
 if the $N$-component  vectors $v_n \equiv (a_{mn})$ are pseudo-orthogonal.

 Now, consider more general nondegenerate matrices $a_{mn}$ and
       define the new scalar fields $x_n$:
\be
       x_n  \equiv  \sum_{m=1}^{N} a_{nm}^{-1} \epsilon_m \psi_m  \, ,
   \,\,\,\,\,\,\,\,\,\,\,\,\,\,\,
       \psi_n  \equiv  \sum_{m=1}^{N} \epsilon_n a_{nm} x_m  \, .
 \label{12}
\ee
 In terms of these fields, Eqs.(\ref{10})  read as
\be
  \p_u \p_v x_m  \equiv  \varepsilon g_m
     \exp {\sum_{k,n=1}^{N}  \epsilon_n a_{nm} a_{nk} x_k}
     \equiv \exp {\sum_{k=1}^{N} A_{mk} x_k}  \, ,
 \label{13}
\ee
 and we see that the symmetric matrix
   \be
  A \equiv a^T \epsilon a \ , \,\,\,\,\,\,\,\, \epsilon_{mn} \equiv
  \epsilon_m \ \delta_{mn} \,,
\label{14}
    \ee
     defines the main properties of the model.

 If $A$ is a diagonal matrix,  we return to the $N$-Liouville model.
 If $A$ is the Cartan matrix of
 a Lie algebra, the system (\ref{13}) coincides with the corresponding
 Toda system, which is integrable and can be more or less explicitly
         solved (see, e.g., \cite{Leznov}, \cite{Saveliev} ).\footnote{
         It can easily be seen that, due to the special structure of
         $a_{mn}$ ($a_{1n} = 1+a_n$, $a_{2n} = 1-a_n$),
         the Cartan matrices of the simple algebras of rank 2 and 3
         cannot be represented in the form (\ref{14}).
         Further analysis shows that this probably is also true
         for any rank. As will be shown in a forthcoming publication,
         any symmetric matrix $A_{mn}$, which is the direct sum of
         a diagonal $L\times L$-matrix $\gamma_n^{-1} \delta_{mn}$
         and of an arbitrary symmetric matrix $\bar{A}_{mn}$,
         can be represented in form (\ref{14}). If $\bar{A}_{mn}$
         is a Cartan matrix, the system (\ref{13}) reduces to $L$
         independent Liouville (Toda $A_1$) equations and the
         higher-rank Toda system.}
 Here we mostly consider
 the $A_N$ Toda systems having very simple solutions. However, the
 solutions have to satisfy the constraints that in terms of $x_n$ are:
 \be
 2\sum_{n=1}^N a_n \ \p_i^2 x_n \ =
 -\sum_{n,m =1}^N   \p_i x_m \ A_{mn} \ \p_i x_n \, , \,\,\,\,\,\,\, i=u,v \, .
 \label{15}
 \ee
 In the $N$-Liouville model the most difficult
 problem was to solve the constraints (\ref{15}) but this problem was
 eventually solved. In the general nonintegrable case
            of an arbitrary matrix
 $A$ we do not know even how to approach this problem. We hope that
 in the Toda case the solution can be somehow derived but this problem is
 not addressed here.
            Instead, in Section~4 we introduce a simplified
 model that can be completely solved.

 Now, let us write the general equations in the form that is particularly
 useful for the Toda systems. Introducing notation
  \be
 X_n \equiv \exp (-\half A_{nn} x_n) \ , \,\,\,\,\,\,
  \Delta_2 (X) \equiv X\ \p_u \p_v X - \p_u X \ \p_v X , \,\,\,\,\,\,
  \alpha_{mn} \equiv -2 A_{mn} / A_{nn} \,,
 \label{16}
 \ee
 it is easy to rewrite Eqs.(\ref{13}) in the form:
 \be
 \Delta_2 (X_n) =
  -\half \varepsilon \ g_n A_{nn} \prod_{m \neq n} X_m^{\alpha_{nm}} \, .
 \label{17}
 \ee
 The multiplier $|-\half \varepsilon \ g_n A_{nn}|$ can be removed by using
 the transformation $x_n \mapsto x_n + \delta_n$ and the final (standard)
 form of the equations of motion is
 \be
 \Delta_2 (X_n) =
  \varepsilon_n  \prod_{m \neq n} X_m^{\alpha_{nm}} \, ,
 \label{18}
 \ee
 where $\varepsilon_n \equiv \pm 1$.

 These equations are in general not integrable. However, when $A_{mn}$
 are the Cartan matrices, they simplify to integrable equations
 (see \cite{Leznov}).
   For example, for the Cartan matrix of $A_N$, only the
   near-diagonal elements of the matrix $\alpha_{mn}$
   are nonvanishing, $\alpha_{n-1,n+1} = 1$.
 This allows one to solve Eq.(\ref{18}) for any $N$.
 The parameters $\alpha_{mn}$ are invariant w.r.t.
 transformations $x_n \mapsto \lambda_n x_n + \delta_n$ and hence
 $A_{mn}$ can be made non-symmetric while preserving
 the standard form of the equations (recall that the Cartan matrices of
 $B_N$, $C_N$, $G_2$, and $F_4$ are not symmetric). In this sense,
 $\alpha_{mn}$ are the fundamental parameters of the equations of
 motion. From this point of view, the characteristic property of the
 Cartan matrices is the simplicity of Eqs.(\ref{18}) which allows one
 to solve them by a generalization of separation of variables.
 As is well known, when $A_{mn}$ is the Cartan matrix of any simple
  algebra, this procedure gives
 the exact general solution (see \cite{Leznov}). In next Section we
 show how to construct the exact general solution for the $A_N$ Toda system
 and write a convenient representation for the general solution that
 differs from the standard one given in \cite{Leznov}.

\section{Solution of the $A_N$ Toda system}

The $A_N$ equations are extremely simple,
 \be
 \Delta_2 (X_n) =
  \varepsilon_n X_{n-1} X_{n+1} \, , \,\,\,\,\,\,\,\,
  X_0 \mapsto 1 \ , \,\,\,\, X_{N+1} \mapsto 1 , \,\,\,\,
 n = 1,...,N ,
 \label{19}
 \ee
 and can be reduced to one equation for $X_1$ by using the relation
 between $\Delta_2 (X)$ and higher determinants, $\Delta_n (X)$
 (see \cite{Leznov}):
 \be
 \Delta_2 (\Delta_n (X)) =
 \Delta_{n-1}(X) \ \Delta_{n+1}(X) \, , \,\,\,\,\,\,\,\,
 \Delta_1(X) \equiv X , \,\,\,\,\,\,  n \geq 2 \, .
 \label{20}
 \ee
 From Eqs.(\ref{19}), (\ref{20}) we find that
 \be
 \Delta_{N+1} (X_1) = \prod_{n} \varepsilon_n \, .
 \label{21}
 \ee
 This equation looks horrible but is known to be soluble.

     Let us start with the Liouville ($A_1$ Toda) equation $\Delta_2(X) = g$
 (see \cite{DPP}, \cite{Gervais}, \cite{Leznov}, \cite{ATF5}).
    Calculating the derivatives of $\Delta_2(X)$ w.r.t. $u$ and $v$,
 we find that
 \be
\label{22}
\p_u^2 \bigl( X^{-1} \, \p_v^2 X \bigr) \,=\,0 \, ,
\qquad
\p_v \bigl( X^{-1} \, \p_u^2 X  \bigr)\,=\,0 \,.
\ee
It follows that if $X$ satisfies (\ref{22})  then there exist some
`potentials' ${\cal{U}}(u)$, ${\cal{V}}(v)$ such that
\be
\label{23}
\p_u^2 X \,-\,{\cal U}(u) \, X\,=\,0 \, , \qquad \p_v^2 X \,-\,{\cal V}(v)
\, X\,=\,0 \,.
\ee
Thus the Liouville solution can be written as (\cite{ATF5})
\be
\label{24}
X(u,v) = \sum a_{\mu}(u) \ C_{\mu \nu} \ b_{\nu}(v) \, ,
\ee
 where $a_{\mu}(u)$ and $b_{\nu}(u)$ ($\mu, \nu = 1,2$) are linearly
 independent solutions of the equations
\be
\label{25}
a''(u) \, - \,{\cal U}(u) \, a(u) \, = \,0,
\qquad
b''(v)  \, - \, {\cal V}(v) \, b(v) \, = \, 0 \, .
\ee
 and $C_{\mu \nu}$ is a nonsingular matrix. As the potentials are
 unknown, the solutions $a_1$, $b_1$ can be taken arbitrary while
 $a_2$, $b_2$ then may be defined by the Wronskian first-order equations
\be
\label{26}
 W[a_1(u), a_2(u)] =1 \ , \,\,\,\,\,\,\,\, W[b_1(v), b_2(v)] =1 \ .
\ee
 The matrix $C_{\mu \nu}$ should obviously satisfy the normalization
condition $\det C = g$.

 We have repeated this well known derivation at some length because it is
 completely applicable to the $A_N$ Toda equation (\ref{21}).
 By similar but rather cumbersome derivations it can be shown that
 $X_1$ satisfy the equations
 \be
 \label{27}
 \p_u^{N+1} X + \sum_{n=0}^{N-1} {\cal U}_n(u) \ \p_u^n X = 0 \, ,
 \qquad
 \p_v^{N+1} X + \sum_{n=0}^{N-1} {\cal V}_n(v) \ \p_v^n X = 0 \, .
 \ee
 Thus the solution of (\ref{21}) can be written in the
 same `separated' form (\ref{24}), where now
 $a_{\mu}(u)$ and $b_{\nu}(v)$ satisfy the ordinary linear differential
 equations of the order $N+1$ (corresponding to Eqs.(\ref{27})),
 with the unit Wronskians,
 \be
 \label{28}
 W[a_1(u),..., a_{N+1}(u)] =1 \ , \,\,\,\,\,\,\,\, W[b_1(v),..., b_{N+1}(v)] =1 \ ,
 \ee
and $\det C = \prod \varepsilon_n$.

 As an exercise, we suggest the reader to prove these statements for $N=2$.
 The key relation that follows from the condition $\p_u \Delta_3 (X) =0$
 is the partial integral
\be
 \label{29}
 \p_v \biggl[ \p_v \biggl({X \over {\p_u X}}\biggr) \, / \,
 \p_v \biggl({{\p_u^3 X} \over {\p_u X}}\biggr) \biggr] = 0\ .
 \ee
 It follows that the expression in the square brackets is equal to
 an arbitrary function $A_0(u)$ and thus we have
 \be
 \label{30}
 \p_v \biggl[ \biggl({X \over {\p_u X}}\biggr) \, + \,
  A_0(u) \, \biggl({{\p_u^3 X} \over {\p_u X}}\biggr) \biggr] = 0\ .
 \ee
 Denoting the expression in the square bracket by $-A_1(u)$ and
 introducing the notation ${\cal U}_1(u) = A_1 / A_0$ and
 ${\cal U}_0(u) = A_1^{-1}$, we get Eq.(\ref{27}) with $N=2$.

 Let us return to the general solution of Eq.(\ref{21}).
 In fact, considering Eqs.(\ref{28}) as inhomogeneous differential
 equations for $a_{N+1}(u)$, $b_{N+1}(v)$ with arbitrary chosen functions
 $a_n(u)$, $b_n(v)$ ($1\leq n \leq  N$), it is easy to write the explicit
 solution of this problem:
 \be
 \label{31}
 a_{N+1}(u) = \sum_{n=1}^N a_n(u) \int^u d\bar u \ W^{-2}_N(\bar u) \
  M_{N,\,  n}(\bar u) \ .
 \ee
 Here $W_N$ is the Wronskian of the arbitrary chosen functions $a_n$ and
 $M_{N, \, n}$ are the complementary minors of the last row in the Wronskian.
 Replacing $a$ by $b$ and $u$ by $v$ we can find the expression for $b_{N+1}(v)$
 from the same formula (\ref{31}). To complete the solution we should
 derive the expressions for all $X_n$ in terms of $a_n$ and $b_n$.
 This can be done with simple combinatorics that allows one to express $X_n$
 in terms of the $n$-th order minors. For example, it is very easy to
 derive the expressions for $X_2$:
  \bdm
 X_2 = \varepsilon_1 \Delta_2 (X_1) =
  \varepsilon_1  \sum_{i<j}  W[a_i(u), a_j(u)] \ W[b_i(v), b_j(v)] \ ,
 \edm
 which is valid for any $N \geq 1$ ($i,j = 1,...,N+1$).
 Note that expressions for all $X_n$
 have a similar separated form. This possibly hints that some rather
 complex separation of variables may give us a tool for (approximate)
 solving more general, nonintegrable equations (\ref{18}).

 Our simple representation of the $A_N$ Toda solution is completely equivalent to
 what one can find in \cite{Leznov} but is more convenient for treating
 some problems. For example, it is useful in discussing asymptotic and
 analytic properties of the solutions of the original physics problems.
 It is especially appropriate for constructing wave-like solutions of
 the Toda system which is similar to the wave solutions of the
 $N$-Liouville model. In fact, quite like the Liouville model,
 the Toda equations
 support the wave-like solutions. To derive them let us first identify
 the moduli space of the Toda solutions. Recalling the $N$-Liouville
 case, we may try to identify the moduli space with the space of the
 potentials ${\cal U}_n(u)$, ${\cal V}_n(v)$. Possibly, this is not
 the best choice and, in fact, in the Liouville case we finally made
 a more cumbersome choice suggested by the solution of the constraints.
 For our present purposes the choice of the potentials is as good as any
 other because each choice of ${\cal U}_n(u)$ and ${\cal V}_n(v)$ defines
 some solution and, vice versa, any solution given by the set of the
 functions ($a_1(u),..., a_{N+1} (u)$), ($b_1 (v),..., b_{N+1} (v)$)
 satisfying
 the Wronskian constraints (\ref{28}) defines the corresponding set of
 the potentials (${\cal U}_0(u ),...,{\cal U}_{N-1}(u)$),
 (${\cal V}_0(v),...,{\cal V}_{N-1}(v)$).

 Now, as in the Liouville case, we may consider the reduction of the
 moduli space to the space of constant `vectors' $(U_0,...,U_{N-1})$,
 $(V_0,...,V_{N-1})$. The fundamental solutions of the equations (\ref{27})
 with these potentials are exponentials (in the nondegenerate case)
 \be
 \label{32}
 a_n (u) = \exp (\mu_n u)\ , \,\,\,\,\,  b_n (v) = \exp (\nu_n v) \, ,
 \,\,\,\,\,\,\,\,
 \sum_{n=1}^{N+1} \mu_n = 0 , \,\,\,\, \sum_{n=1}^{N+1} \nu_n = 0 .
  \ee
 In this reduced case we may regard the space of the parameters
 ($\mu_n$, $\nu_n$) the new moduli space, in complete agreement with
 the Liouville case. Of course, the constraints (\ref{11}) define
 further restrictions on the  moduli ($\mu_n$, $\nu_n$) but here we do
 not address this problem.

 Note only that, as in the $N$-Liouville case, one can construct
 nonsingular waves. To show this is not much more difficult than
 in the Liouville case but requires more lengthy derivations.
 We hope to publish these in a forthcoming paper.

\section{A simple integrable model of (1+1)-dimensional dilaton
gravity coupled to Toda scalar matter}

 Let us suppose that the potential $V$ is independent of $\f$, i.e.
 $V(\f, \psi) \equiv V(\psi)$\footnote{In this model we suppose that
 there are $N$ scalar matter fields $\psi_n$ with $n=1,...,N$ while
 $F$ is trivial and $\f$ is treated separately}.
 Then Eq.(\ref{F.18}) is simply the D'Alembert equation for $F(u,v)$. It
 follows that the metric can be written as
 \bdm
 f= \ve a^{\prime}(u) \ b^{\prime}(v).
 \edm
 Due to the residual freedom of the
 coordinate choice in the light-cone metric we can choose $(a,b)$ as the
 new (local) coordinates and then denote them by $(u,v)$.
 In this coordinates and notation we simply have $f=\ve$ and $F=0$
 in all the equations. We thus see that the equations (\ref{F.16})   are
 independent of $\f$ and can be solved independently  of the equations
 (\ref{F.15}), (\ref{F.17}). Suppose that the matter fields $\psi$ are known
 and first solve equations (\ref{F.15}) and (\ref{F.17}).

 The general solution of Eq.(\ref{F.15}) can be written as
\be
 \label{33}
 \f = -\ve \int_0^u \int_0^v d\bu d\bv \ V[\psi (\bu , \bv)] +
 A(u) + B(v) ,
\ee
where $A(u)$, $B(v)$ are arbitrary functions. The constraints (\ref{F.17})
in this model have the form
\be
 \p_i^2 \f  = -\sum_{n=1}^N  (\p_i \psi_n)^2 , \,\,\,\,\,\,\, i=u,v \, .
 \label{34}
\ee
Using (\ref{F.16}) we easily derive
\be
 \p_i V  = \ve \p_j \sum_1^N (\p_i \psi_n)^2 ,
 \,\,\,\,\,\,\, (i,j)=(u,v) \,\, {\rm or} \,\, (v,u) \,,
 \label{35}
\ee
find $A(a)$, $B(b)$ in terms $\psi$, and finally obtain:
\be
 \label{34}
 \f = -\ve \int_0^u \int_0^v d\bu d\bv \ V[\psi (\bu , \bv)] -
 \int_0^u d\bar{\bu}  \int_0^{\bar{\bu}}  d\bu \ \Phi_u (\bu) -
 \int_0^v d\bar{\bv}  \int_0^{\bar{\bv}}  d\bv \ \Phi_v (\bv) +
  A^{\prime}(0) u + B^{\prime}(0) v \,,
\ee
where we omitted the unimportant arbitrary term $A(0) + B(0) = \f (0,0)$
and denoted
\bdm
 \Phi_u (u) \equiv \sum_1^N (\p_u \psi_n (u,0))^2 \,, \,\,\,\,\,\,\,\,\,\,\,
 \Phi_v (v) \equiv \sum_1^N (\p_v \psi_n (v,0))^2 \,.
\edm

      Now, to get integrable equations for $\psi$ we take the potential
 (\ref{8}) with $q_n^{(0)}$ given by the r.h.s. of Eq.(\ref{9a}).
 Then, we can use for the scalar fields the  equations
 (\ref{10}) and (\ref{12}) -- (\ref{14}).
        If we take the potential for which the $\psi$ equations of
 motion can be reduced to integrable Toda equations we find an explicit
 solution for the nontrivial class of dilaton gravity minimally
 coupled to scalar matter fields. This model is a very complex
 generalization of the well studied CGHS model in which the scalar fields are
 free and $V=g$. In future, we plan a detailed study of the $A_N$ case.
 The easiest case is $N=1$ (the Liouville equation for one $\psi$). The
 first really interesting but simple theory is the case of two scalar fields
 satisfying the $A_2$ Toda equations. Taking, for example,
 \bdm
 V = \exp{(\sqrt 3 \ \psi_1 - \psi_2)} + \exp{(2 \ \psi_2 )} ,
 \edm
 we find the simplest realization of the $A_2$ Toda dilaton gravity model
      the complete solution of which can be obtained by use of the above
      derivations.

 As a simple exercise one may consider the
  reduction from dimension (1+1) both to the dimension (1+0)
 (`cosmological' reduction) and to the dimension (0+1)
 (`static' or `black hole' reduction)
 as well as the moduli space reduction to waves. One of  the most interesting
 problems for future investigations is the connection between these three
 objects. It was discovered in the $N$-Liouville theory but now  we see
        that it can be found in a much more complex theory described by
       the Toda equations. It is not impossible that the connection also
       exists  (in a weaker form?) in some nonintegrable theories, say, in
 theories close to the Toda models.

 Note in conclusion, that the one-dimensional Toda
 equations  were earlier employed mostly in
 connection with the cosmological and black hole solutions
 (see, e.g. \cite{Fre} - \cite{Sorin}). To include into consideration
 the waves one has to step up at least on dimension higher. The principal
 aim of the present paper was to make the first step and
 explore this problem in a simplest
 two-dimensional Toda environment.


 A more detailed  presentation will be published elsewhere.


\bigskip
\bigskip

 {\bf Acknowledgment:}

 The author appreciates financial support from the Department of Theoretical
 Physics of the University of Turin and INFN (Turin Section) and of the Theory
 Division of CERN, where some results were obtained.
 The useful discussions with V.~de Alfaro and A.~Sorin are kindly acknowledged.
 The author is especially grateful to V.~de Alfaro for his support during
 many years and very fruitful collaboration.

 This work was supported in part by the Russian Foundation for Basic
 Research (Grant No. 06-01-00627-a).


\end{document}